\documentstyle[12pt]{article}

\def \be  {\begin{equation}}
\def \ee  {\end{equation}}
\def \ba  {\begin{eqnarray}}
\def \ea  {\end{eqnarray}}
\begin{document}

\begin{flushright}
ITP-SB-98-45
\end{flushright}

\begin{center}
{\Large \bf Resummation, Power Corrections and 
Prediction in Perturbative QCD\footnote{This preprint
summarizes two presentations at {\it QCD and High Energy Hadronic Interactions}, XXXIIIrd
Rencontres de Moriond, Les Arcs, Savoie, France, 21-28 March, 1998.}} \\

\bigskip 

 George Sterman \\
{\it Institute for Theoretical Physics,
         State University of New York at Stony Brook, \\  
         Stony Brook,  NY 11794-3840}

\end{center}

\begin{abstract}
I give a pedagogical introduction to
resummation and power corrections, using the thrust
variable in electron-positron annihilation as an example, followed by an discussion
of issues of predictability in perturbative QCD.
\end{abstract}

\begin{center}
(June, 1998)
\end{center}

\section{The Thrust and Time Evolution}

The session on the predictive power of QCD emphasized
new developments in power corrections and resummation.
This section presents material that
introduced the session, with a discussion of some of 
the basic ideas and methods that underly recent progress \cite{reviews,irr,kostnp}
on these topics, using the familiar example
of the thrust in ${\rm e}^+{\rm e}^-$ annihilation.  The following sections briefly
treat some broad issues of predictive power.  These comments were followed by an open
discussion at the session.

{\it Thrust.} The thrust in ${\rm e}^+{\rm e}^-$ annihilation may be
defined as
\be
T={\rm max}_{\hat{n}}\, \sum_{{\rm particles}\, i}
{|\vec{p}_i||\cos\theta_i| \over Q}\, ,
\ee
where the sum is over all particles in  the final state, and
where $\cos\theta_i$ is the angle  between momentum  $\vec p_i$
and the axis defined by the vector, $\hat n$, which is chosen
to  maximize the sum in the ${\rm e}^+{\rm e}^-$ center-of-mass
(c.m.) frame.    Neglecting particle masses, and taking
$Q$ to be the total c.m.\ energy, the maximum  value of
$T$ is unity.  In this configuration, the final state consists of
two opposite-moving jets, $J_1$ and $J_2$, 
each consisting of perfectly collinear
particles. 
For definiteness, we take the direction of $J_1$ along $\hat n$,
with $J_2$ opposite.
The probability for such a final state is suppressed
by radiation, and it is this suppression, associated with the
long-time evolution of the system, which we study using resummation
methods.  Let us see how this works.

We first observe that for massless particles
$|\vec p_i|=E_i$, so that $\sum |\vec p_i|=Q$.  
Then, defining $p_i^\pm=E_i \pm |\vec{p}_i||\cos\theta_i|$,
we find
\be
1-T = \sum_{i\in J_1} {p_i^-\over Q} + \sum_{j\in J_2} {p_j^+\over Q}
\sim \sum_{i\in J_1} {1\over \tau_i^+ Q} + \sum_{j\in J_2} {1\over \tau_j^- Q}\, ,
\label{1-t}
\ee
where, invoking the uncertainty principle, we identify $\tau_i^\pm\sim 1/p_i^\mp$
as the space-time variable conjugate to the momentum component $p_i^\pm$.
Elementary relativistic kinematics then shows that $\tau_i^\pm$ is the
typical time that it takes to emit a particle of momentum $p_i$, as seen
in the c.m.\ frame.  The sums in Eq.\ (\ref{1-t}) are dominated by 
terms corresponding to the ``earliest" emission, for which $\tau^\pm$ is smallest.

{\it Factorization.} From the above, we
can conclude that for small values of $1-T$, the evolution of the quark pair into the final
state is a relatively long-time process.  In this case, it is natural 
to propose that the overall cross section is a product of functions, one
for each jet \cite{thrustresum},
\begin{equation}
{d\sigma\over dT_1\; dT_2}
=J_1(1-T_1)\; J_2(1-T_2)\, ,
\label{sigfact}
\ee
with $(1/2)(1-T_{1,2})\equiv\sum | p_i^\pm/Q|$.  To find $d\sigma/dT$,
we only need to integrate over $T_1$ and $T_2$ in (\ref{sigfact}), subject
to $T=(1/2)(T_1+T_2)$.  

Now consider the significance of the observation that the $T_i$-dependence of
jet $J_i$ is determined by its earliest emission.  We use one of the basic features
of QCD (indeed of quantum field theory) -- that processes occuring at different time scales are 
quantum-mechanically ``incoherent", and may therefore be treated 
probabilistically \cite{incoh,dglap,cssrv}.
In this language, we write
\be
J_1(T_1)
=
P\left(1-T_1=2k^-/Q\right)\; \tilde{J}_0(k^-/Q)\, ,
\label{JtildeJ}
\ee
where $P\left(1-T\right)$ is the probability density for the
emission of a  gluon with $k^-=(1-T_1)Q/2$, while $\tilde{J}_0(k^-/Q)$
is the probability that there has been no emission of gluons before
time scale $\tau^+\sim 1/k^-\sim 1/(1-T_1)Q$. Evidently, $\tilde{J}_0$ is
a function that decreases with time, according to exactly the probability
density $P$,
\be
{dJ_0(k^-/Q)\over d(1-T)}=
\left({Q\over 2}\right)\; {d\tilde{J}_0(k^-/Q)\over dk^-}=-P(2k^-/Q)\tilde{J}_0(k^-/Q)\, .
\label{tildeJevol}
\ee
Now $P(2k^-/Q)$ is just the square of the amplitude for
a quark of energy $Q/2$,
moving in the $\hat n$-direction to emit a gluon with minus momentum $k^-$.  Keeping only the
leading logarithm in $1-T\sim 2k^-/Q$, we find from a lowest-order
calculation that
\be
P\left( 1-T \right ) = C_F\; {\alpha_s\over \pi}\; {\ln (1-T)^{-1}\over 1-T}\, ,
\label{pll}
\ee
where $C_F=4/3$ in QCD.  Substituting (\ref{pll}) into (\ref{tildeJevol})
we find that $J_0$ is a rapidly decreasing function of $1-T$,
\be
\tilde{J}_0(1-T_1)=
\exp \left[ -C_F{\alpha_s\over 2\pi}\ln^2(1-T_1)\right]\, .
\label{tilJsoln}
\ee
Then for $J_1$, and similarly for $J_2$, we have 
\be
{J}_1(1-T_1)=
C_F{\alpha_s\over\pi}\; {\ln(1-T_1)\over 1-T_1}\;
\exp \left[ -C_F{\alpha_s\over 2 \pi}\ln^2(1-T_1)\right]\, .
\label{Jsoln}
\ee
Convoluting the two jets together, we find that the cross section $d\sigma/dT$
takes on a very similar form \cite{thrustresum}, in terms of $1-T$.

{\it Enter the running coupling.}  At this point we ask,
``what about the running of $\alpha_s$?"  This effect enters
through the calculation of the probability density $P(1-T)$,
which we recall is simply a branching
probability for the emission of a single gluon of minus
momentum $(1-T)Q/2$.  As an integral in $k$-phase space this is,
\be
P(1-T) = {2C_F\over 1-T}{\alpha_s\over\pi}\; \int_{(1-T)^2Q^2}^{(1-T)Q^2}\; {dk_T'{}^2 \over k_T'{}^2}\, .
\label{pint}
\ee
The running coupling organizes quantum corrections to this
process that come from momentum scales larger than $k_T$,
the momentum transfer in the emission process.  The natural
extension of (\ref{pint}) to include these effects is thus \cite{run},
\be
P(1-T) \rightarrow {C_F\over 1-T}\; 
\int_{(1-T)^2Q^2}^{(1-T)Q^2}\; {dk_T'{}^2 \over k_T'{}^2}\;
{\alpha_s(k_T'{}^2)\over\pi}\, ,
\label{prc}
\ee
where the running coupling is the familiar
$\alpha_s(k_T^2)=4\pi/\beta_2\ln(k_T^2/\Lambda^2)$, with
$\beta_2=11-2n_f/3$.  
When $1-T=\Lambda/Q$, this expression develops
a singularity, as the perturbative running coupling diverges. 
This is not surprising, since in this region, we are
looking at the emission of very soft gluons.  
The divergence is itself unphysical, but
signals the entry of long-distance physics into
the problem.   The full cross section behaves in much
the same way.

Fuller treatments, which include nonleading logarithms, 
virtual corrections and which
pay closer attention to momentum conservation, show 
the same pattern in the thrust and other 
observables involving jets \cite{thrustresum,gscls}.  Resummed perturbative cross sections almost
always  encounter a ``corner" of phase space where the coupling
blows up \cite{reviews,irr,kostnp}.  At the same time, such cross sections also
often have a surprising feature, illustrated by
Eq.\ (\ref{prc}).  Suppose we regularize the $k_T$ integral
in (\ref{prc}) by adding a mass to $k_T$: $k_T^2\rightarrow k_T^2+m^2$,
with $m>\Lambda$.
Then the regulated coefficient of $1/(1-T)$ in $P(1-T,m)$
is finite all the way down to $T=1$.  For large, fixed $1-T$, however,
$P(1-T,m)$ can be expanded as a power series in $m/Q$,
starting with $P(1-T,m=0)$. 
For $(1-T)Q\gg \Lambda$, the effect of infrared regulation is
to require power corrections to the perturbative result.
``Freezing" $\alpha_s(k_T)$
at some fixed $\alpha_0$ for small $k_T$ has much the same effect.

These observations are quite general; resummation almost always
implies the presence of power corrections in terms of the relevant hard
scale.  It is easy to check, that in the 
range of a few to ten GeV, the quantity $\Lambda/Q$ is between the
``leading order" quantity $1/\ln(Q^2/\Lambda^2)\sim \alpha_s(Q^2)$ and the
``next-to-leading order", $1/\ln^2(Q^2/\Lambda^2)\sim\alpha_s^2(Q^2)$
in magnitude, while
in the same range $\Lambda^2/Q^2$ lies between next-to-leading
and next-to-next-to-leading order corrections.  

We may consider these power corrections as simply a signal of
the limitations of perturbative methods.  Alternately, we
may treat them as a reflection of the structure of 
the actual behavior the full theory \cite{reviews,irr,kostnp}.  This viewpoint is explored
in a number of the talks that followed in the session on the predictive
power of QCD \cite{talks}.

\section{The Predictive Power of Perturbative QCD: A Sketch}

\subsection{Review of basic methods}

{\it Infrared Safety.} The quantitative predictions of perturbative QCD all
require the identification of ``infrared safe" quantities, which 
are dominated by the short-distance, partonic behavior of the theory.
The perturbative expansion for such a quantity,  labelled generically $\sigma_{\rm IRS}$,
depending on a single hard scale, $Q$,  and  some set of dimensionless parameters $C_i$
is
\begin{equation}
\sigma_{\rm IRS}\left({Q^2\over\mu^2},\alpha_s(\mu^2),C_i\right)
=
\sum_{n\ge 0} a_n\left({Q^2\over\mu^2},C_i\right)\; \left[{\alpha_s(\mu^2)\over\pi}\right]^2\, ,
\label{IRSsum}
\ee
where the $a_n$ are dimensionless coefficients.
Assuming that $\sigma_{\rm IRS}$ is observable, it is
independent of the renormalization scale $\mu$.  We may then pick $\mu=Q$, and
get an expansion in the ``small" coupling $\alpha_s(Q^2)$.  Here, the asymptotic
freedom of QCD, by which $\alpha_s$ decreases as its momentum scale increases,
plays a central role.  Note that $\sigma_{\rm IRS}$ need not be a cross section;
we shall discuss other examples below.   The first few $a_n$ are known
for many quantities.  What Eq.\ (\ref{IRSsum}) leaves out, however, are
``power" corrections, that introduce nonperturbative scales.  We have
seen above an  example of how such corrections 
arise, and have seen how perturbation theory
can, at least in some cases, provide guidance as to their nature.  To understand
power corrections in a broader context, it is useful to review how results of
the form (\ref{IRSsum}) are derived.

Perhaps the purest example of this procedure is $\sigma_{\rm tot}^{{\rm e}^+{\rm e}^-}$,
the total cross section for electron-positron annihilation into hadrons.  
In this case, the IR safety of the perturbative sum is ensured by a very fundamental
property of the theory, its unitarity.  Schematically, the optical theorem
(a direct consequence of the conservation of probability) implies that
\be
\sigma_{\rm tot}^{{\rm e}^+{\rm e}^-}(Q)
=
{1\over Q^4}\; {\rm Im}\; \Pi(Q^2)\, ,
\label{sigtot}
\ee
where $\Pi(Q^2)$ is the contribution of all hadronic virtual
states to the forward-scattering amplitude of a single off-shell photon (or Z).
This forward-scattering amplitude is of the general form $\int d^4x \exp[-iq\cdot x]\,
\langle 0|J^\mu(0)\; J_\mu(x)|0\rangle$, with $J$ an electroweak
current (including the electron charge)
 and $q^2=Q^2$.  Such a vacuum matrix element may be treated by the operator
product expansion (OPE), an observation that lies at the heart of the many successes
of the method of QCD sum rules.  The OPE predicts in this case
 that nonperturbative corrections
(proportional to vacuum condensates) begin to contribute only at the level of
$Q^{-4}$ compared to the leading perturbative expansion, (\ref{IRSsum}).  
In fact, this  result can also be derived by a variant of the reasoning given
in Sec.\ 1 above. Perturbatively, power corrections appear through the coupling
of soft gluons, with momenta of order $\Lambda_{\rm QCD}$, to
off-shell partons.  Mueller \cite{mu85}, applying observations
of 't Hooft \cite{tHooft} showed long ago how the running coupling, and
consequently divergences in the perturbative sum, appear from such configurations,
with exactly the power behavior implied by the OPE.  

The direct applicability of the OPE to an IR safe cross section is the 
exception rather than the rule, however, because it is rare to be able to
reduce a physical cross section to a simple product of local operators.  
IR safe quantities for which the OPE does not
control power corrections include event shapes in ${\rm e}^+{\rm e}^-$
annihilation, of which the thrust, discussed above, is a prime example.
The proof of the IR safety of event shapes again depends upon the unitarity
of QCD, but in a modified form \cite{gs78}.  Because different regions of the
final-state phase space are weighted differently in an event shape, 
we cannot apply the optical theorem to the entire process.
Essentially, unitarity is applied separately to each ``jet" of outgoing
partons moving in the same direction.  The resulting sum is indeed IR safe, but the low
momentum scales enter now  through the couplings of soft gluons to
lines that are generally close to the mass shell, as in the example of
thrust above.  Unprotected by the OPE, event shapes inherit, as above,
larger power corrections than the total annihilation cross section, suppressed
by lower powers of $Q$.

{\it Factorization and evolution.}  
The predictive potential of perturbative QCD is greatly enhanced by the
factorization \cite{cssrv}  of short-distance (perturbative) and long-distance 
(nonperturbative) dynamics,
 and by the
computable evolution of cross sections with
momentum transfer.   Examples include
the factorized structure functions in the deeply inelastic scattering (DIS)
of hadron $A$ by vector boson $V$, with spacelike virtuality $q^2=-Q^2$.
These functions take the form
\be
F^V_A\left(x,Q^2\right)
=
\sum_{{\rm partons}\ a}\int_x^1 {d\xi \over \xi}\; 
C_a^V\left({x\over\xi},{Q^2\over\mu^2},\alpha_s(\mu^2)\right)\, \phi_{a/A}\left(\xi,\mu^2\right)
\equiv \sum_a C^V_a\otimes 
\phi_{a/A}\, .
\label{Ffact}
\ee
The $\phi_{a/A}$ are nonperturbative parton
distributions for parton $a$ in hadron $A$, 
while each $C_a^V$ is a series in $\alpha_s(\mu)$, whose lowest order
approximation is found from the Born cross section for $V(Q)+a$ scattering.  
$C_a^V$ is infrared safe, and therefore calculable in perturbation
theory.  The scale $\mu$ is called the factorization scale.  
A key observation is that $C_a^V$ depends on $\mu$ only through
$Q/\mu$, and that all of the $Q$-dependence in $F$ is in  the $C$'s.
The convolution in (\ref{Ffact}) must be independent of the arbitrary choice
of $\mu$, which leads, by simple separation of variable arguments, to
the DGLAP evolution equations \cite{dglap},
\ba
\mu{dF^V_A \over d\mu}&=&0 \quad \quad  \Rightarrow \nonumber\\
\mu{dC^V_a\over d\mu}=\sum_b C^V_b\otimes P_{ba}\left(\alpha_s(\mu^2)\right)\, &\ &\
\mu{d\phi_{a/A}\over d\mu}=-\sum_c P_{ac}\left(\alpha_s(\mu^2)\right)\otimes \phi_{c/A}\, .
\label{dglap}
\ea
The splitting functions $P_{ba}(\eta,\alpha_s(\mu))$ play the role of ``separation
constants", which can depend only on parameters held in common by the parton
distributions $\phi$ and the coefficient functions $C$.

Like the infrared safety of the 
total annihilation cross section, the 
factorization of the inclusive DIS structure functions follows from
the optical theorem, and may be treated by means of the operator
product expansion.  This opens the door for particularly
well-organized treatments of power corrections, some of which were discussed in this session
\cite{disIRR}.

Closely related, but not covered by the OPE directly, are the factorizability and evolution
equations of fragmentation functions,
\be
\mu{dD_{C/c}(z,\mu)\over d\mu}
=
\sum_d
\int_z^1 {dy\over y}\; {\alpha_s\over \pi}\; P_{cd}(y)\; D_{C/d}\left({z\over y},\mu\right)\, .
\label{Devol}
\ee
A generic perturbative QCD cross section is factorized schematically as
\be
\sigma_{AB\rightarrow C} = 
\sum_{abc}\; \phi_{a/A} \otimes \phi_{b/B} \otimes H_{abc} 
      \otimes D_{C/c}\, ,
\label{generic}
\ee
where in the case of a jet final state, the 
analog of the fragmentation function $D_{C/c}$
is itself perturbatively calculable, and can be absorbed
into $H_{abc}$.   This factorization reflects the mutual
independence of QCD dynamics at different length scales, and expresses
the same physical principles that are exploited in methods based on
``effective theories" for heavy quark physics.

\subsection{Successes and limitations}

The evolution of structure functions is probably the most ``precise" and ``predictive"
of perturbative QCD results, tying together a multitude of experiments over a wide range
of momentum transfers \cite{evolexp}.  In this case, the cross section is nearly the 
simplest form possible in  (\ref{generic}), involving the parton distributions
of only a single hadron. Other striking phenomenological successes include the somewhat
more qualitative predictions based on improvements of fragmentation analysis
using on the concept of ``coherence" \cite{cohere}.

In the experimental presentations during the first half of this conference, we have also
seen:

\begin{itemize}
\item{} Good, sometimes outstanding, but not universal success
(at the level of tens of percent) for jet and direct photon data
over varying momentum transfers and magnitudes of cross sections {\protect\cite{jetpresent}}.

\item{} Estimates of theoretical uncertainties that more often
than not exceed the data errors.

\item{} Uneven success with heavy flavor production at high energies {\protect\cite{hqpresent}}.

\end{itemize}

From the point of view of Eq.\ (\ref{generic}), the 
reasons why the
predictions of perturbative QCD are sometimes
 of limited accuracy are fairly clear:  

\begin{itemize}
\item{} The hard scattering functions $H_{abc}$ in Eq.\ (\ref{generic})
have generally been computed to NLO in $\alpha_s(\mu)$.  Their variation
with $\mu$ is of the next highest order, $\mu dH_{abc}/d\mu \sim \alpha_s(\mu)^{\rm NLO+1}$.

\item{} Each nonperturbative function in Eq.\ (\ref{generic}) requires a
factorization scale $\mu_f$.  Although simplicity suggests that all of
the $\mu_f$ be chosen equal, a single choice may not be the ``optimal" for
all, which can affect the size of higher-order corrections in the hard scattering
function.

\item{} For complex processes, there are many ``implicit" dimensionless scales.
Examples include the $R$-parameters used to define jets in ${\rm p}\bar{\rm p}$
collisions.  Dependence on such parameters is at once 
potentially important and hard to quantify systematically \cite{seymour}.

\item{} The ratios of partonic energies to heavy quark masses are often large,
giving additional dimensionless parameters.

\item{} Parton distribution and fragmentation functions must be extracted from experiment,
and include uncertainties that depend on choices of which data to use,
and on starting parameterizations \cite{keller}.

\item{} Power corrections of all kinds can be important, as we have
suggested above.
For event shapes in electron-positron annihilation there is ample evidence of
this, and in this session experimental evidence and
theoretical progress along these lines was discussed.
In hadronic scattering at collider energies, energy flow from the ``underlying event"
is another potentially significant power correction \cite{seymour}.  
``Intrinsic" transverse momentum is yet another \cite{yuan}.
It is the nature of power corrections to depend sensitively on kinematics, and they
can be crucial in some kinematic ranges, and negligible not far away in phase space.
\end{itemize}

All of these limitations, particularly those
associated with power corrections, sound a note of caution for our interpretation
of NLO predictions based upon perturbative QCD.  Let me make what I consider an important comment,
however, in this connection.  If we are
interested in using existing perturbative 
predictions for new-particle production, or for measuring $\alpha_s$,
power corrections are something of an embarresment, and we should emphasize
quantities for which they are minimized.  If, on the other hand, we are
interested in the dynamics of QCD at the perturbative-nonperturbative
interface, then power corrections are a boon.  As we have observed above, at scales
of a few GeV, power corrections are typically smaller than the leading order, but competitive with
next-to-leading order.  They are therefore readily observable 
but not dominant, and teach
us new things about quantum chromodynamics.  As with most new information,
we must develop the tools with which to interpret them.
Let me turn briefly to some preliminary suggestions on how this might be done.

\section{The Operator Content of Nonperturbative Parameters}

If we had not known of the local gluon condensate from the general
considerations of the OPE, we might have stumbled upon it in the course of
analyzing infrared renormalons in the total electron-positron annihilation
cross section.
It is natural to expect that infrared renormalons associated
with event shapes and related semi-inclusive cross sections  
might imply a phenomenological role for as-yet unappreciated
nonperturbative quantities,  which, like the gluon condensate,
are expressible in terms of the expectation values of operators
in QCD.  The literature routinely refers to ``generalizations"
of the OPE.  Whether such generalizations constitute real progress
remains to be seen, of course, but let me give an example or
two, which illustrate a few of the possibilities.

{\it The $Q_T$-distribution and ``Wilson lines".}
That resummation implies nonperturbative corrections is not a new observation.
In fact, there is a quite sophisticated formalism,
that predates the wave of interest in infrared renormalons,
 in  at least one case of considerable
phenomenological interest, the transverse momentum distributions for Drell-Yan
pairs \cite{yuan,dyother,csdy}.   In this cross section,  the resummation is most transparent in Fourier transform
(``impact parameter") space, where it exponentiates into
a form  that reminds us of the thrust distribution above,
\begin{eqnarray}
\tilde \sigma(b)
&\sim& \exp \left[
\int_0^Q {d^2k_T\over k_T^2}\; A(\alpha_s \left(k_T^2)\right)\; \ln\left({Q^2\over k_T^2}\right)\;
\left({\rm e}^{-ik_T\cdot b}-1\right) + \dots \right]
\nonumber\\
&\sim& \exp \left[ \int_{1/b_*^2}^Q {d^2k_T\over k_T^2}\; A(\alpha_s \left(k_T^2)\right)\; 
\ln\left({Q^2\over k_T^2}\right)
+ g_2(b_0) b^2\ln \left(Qb\right) + \dots \right]\, .
\label{qtsum}
\end{eqnarray}
The function $A=2C_F(\alpha_s/\pi)+\dots$ is a power series in the strong coupling
with finite coefficients, so that all logarithms  of the impact parameter, $b$, are
generated by the explicit
integrals in Eq.\ (\ref{qtsum}) and by expansions of the running coupling.  
The running coupling, however, diverges for $k_T\sim \Lambda_{\rm QCD}$
in the first line.  In  the second expression, the $k_T$ integral has been regularized,
in such a way that all  leading (and in the full form) next-to-leading logarithms
in $b$ are retained.  This is done by introducing a modified impact
 parameter $b_*=b/\sqrt{1+b^2/b_0^2}$,
with $b_0$ a fixed distance scale \cite{csdy}.  The ``renormalon" singularity at small $k_T$ is
now taken  care of, but at the price of introducing a new mass scale,
$1/b_0$, in the problem.
This, however, is just what we  expect for a full perturbative-nonperturbative
cross section.  In this case, the analog of the power expansion in  $1/Q$ above is
an expansion in powers of $b$.  
Of special interest is  the ``$g_2$" term shown in the second form in Eq.\ (\ref{qtsum}), in which
a perturbative logarithmic dependence on $Q$ multiplies a nonperturbative power
of $b$.  In principle, it is possible to measure these, and  related nonperturbative
parameters.  The derived values will depend on the choice of $b_0$.  
Considerable attention has been given to this problem recently in
the context of electroweak vector boson production at the Tevatron \cite{yuan,Wtalks,kellis}.

Turning to the new parameter $g_2(b_0)$;  what might it be telling us?
One  approach, which is close in spirit to the OPE analysis for annihilation,
is to look for an operator vacuum expectation value which, when expanded
in perturbation theory, gives an expression that is the same as the
low-$k_T$ tail of the integral  in Eq.\ (\ref{qtsum}) when the
exponential in the first line is expanded to order $b^2$.   Such an operator
exists,  and is, in fact, a reasonably natural generalization of
the gluon condensate in the OPE.  We can build up this object out of
the elementary operators of QCD \cite{kostnp}.  

We begin by introducing a ``Wilson line", or ordered exponential of the
gluon field, along a light-like path,
\begin{equation}
\Phi_v(x,x+sv)=P\exp \left[ -ig\int_0^s dt\; v\cdot A(x+tv)\right]\, .
\label{Phidef}
\end{equation}
Here $P$ stands for ``path-ordering", which simply means that
we keep track of the color indices of each field $A^\mu$
in the expansion on (\ref{Phidef}).
In $\Phi_v$, the gluon field is integrated along a straight path
in space-time, defined by the ``velocity" vector $v^\mu$, which begins
at point $x$, and ends at point $sv^\mu+x^\mu$.   We can
think of such an operator as a  model for the interaction
of the gauge field $A^\mu$ with a fast-moving quark of
velocity $v^\mu$, neglecting recoil.  
The analogous quantity for a fast-moving electron
interacting with the photon field is a pure phase,
and (\ref{Phidef}) is sometimes also referred to as 
the quark's ``nonabelian phase".

The influence of soft gluons on the amplitude for the
annihilation of a quark-antiquark pair can be generated perturbatively
by sewing together two Wilson lines, one representing the quark,
the other the antiquark,
\begin{equation}
U_{v_1v_2}(0)=T\left[ \Phi_{v_2}^\dagger(0,-\infty)\Phi_{v_1}(0,-\infty) \right]\, ,
\label{Wdef}
\end{equation}
with $T$ the time-orderding operation.
We are interested in the transverse momentum of radiated gluons, which
in classical terms is associated with the transverse force experienced
by the quarks before their annihilation.  Recalling the Lorentz force
of electrodynamics, a gauge invariant measure of the nonabelian Lorentz
force is
\begin{equation}
{\cal F}_v^\alpha(x)
=
-ig \int_{-\infty}^0
ds\; \Phi_v(x,x+sv)\; v_\mu F^{\mu\alpha}(sv+x)\; \Phi_{-v}(x+sv,x)\, ,
\label{calFdef}
\end{equation}
which we insert along the paths of the Wilson lines in $U$, to get
\begin{equation}
\langle 0|\; \big | \Phi^\dagger_{v_2}(0,-\infty)
\left( \vec {\cal F}_{v_1}(0) - \vec {\cal F}_{v_2}^\dagger(0) \right) 
\Phi_{v_1}(0,-\infty) \big |^2\; 
|0\rangle\, .
\label{nlocal}
\end{equation}
This rather nontrivial generalization of the gluon condensate,
$\langle 0|F^{\mu\nu}(0)F_{\mu\nu}(0)|0\rangle$, 
generates the soft tail of the $k_T$ integral in Eq.\ (\ref{qtsum}),
as desired.   It is a matrix element that is sufficiently general
to be found in many phenomenologically-interesting contexts.
It is certainly a relevant question whether this ``universality"
can be put to good use.

Although (\ref{nlocal}) is nonlocal, it is contained along a 
one-dimensional path.  This relative simplicity is 
associated with measuring a total momentum component
of all the hadrons in the final state, in this case the total transverse
momentum of hadrons recoiling against an electroweak boson.  Other quantities, like the thrust,
which measure details of the final state, are associated with even more complex
operators \cite{Tk,KorOS}.  For example, power corrections to the  thrust are sensitive to
final-state interactions, in a way that the total transverse momentum is not.
Power corrections to the thrust are associated with operators on
the sphere at infinity, which is reached  only after even the softest
of interactions has ceased.  Schematically, the first power correction
to the
moments of the
thrust, $\tilde\sigma(N)\equiv\int dTT^Nd\sigma/dT$,   may be represented as \cite{KorOS}
\begin{equation}
\ln {\tilde \sigma(N)} \sim {\rm PT} +{N\over Q}\int_{-1}^1 d\cos\theta \left( 1-|\cos\theta|\right) 
{\cal E}(\cos\theta)
\label{npthrust}
\end{equation}
where ``PT" denotes a perturbative contribution,
and where $\cal  E$ is a matrix element that measures
energy flow in the presence of a pair  of Wilson
lines, which represent the incoming quarks.  To be specific,
\begin{equation}
{\cal E}(\cos\theta)
=
\langle 0|\; U^\dagger_{v_1v_2}(0)\; \Theta(\cos\theta)\; U_{v_1v_2}(0)\; |0\rangle\, ,
\label{calEdef}
\end{equation}
where $U$ is given by Eq.\ (\ref{Wdef}),
and where $\Theta$ is related to the energy-momentum tensor $\theta_{\mu\nu}(y)$ 
at infinity \cite{Tk,KorOS},
\begin{equation}
\Theta(\cos\theta)
=
\lim_{|\vec y|\rightarrow\infty}\;
\int_0^\infty dy_0\; d\Omega'_{ij}(\vec y)\epsilon_{ijk}\theta_{0k}(y^\mu)\;
\delta(\cos\theta'-\cos\theta)\, .
\label{EM}
\end{equation}
Such ``maximally-nonlocal" correlation functions seem to be necessary to
describe the nonperturbative information in power corrections to  thrust.
Again, more thought will be needed to test the utility of this observation.

\section{Comments and Conclusions}  
  
Thanks to the program of computing a wide variety of hard-scattering
cross sections at next-to-leading order \cite{glover}, we can pose more questions
of perturbative QCD than ever before.  We must recognize, however,
that NLO is the first serious approximation, and is not guaranteed
to work, unaided, except in cases where there is only a single
large scale.  Beyond these situations, we may have to resort to
  higher orders, resummation and/or power corrections.
Despite these limitations, pQCD at NLO has led to striking successes \cite{glover}, even
compared to a few years ago, many of which have been shown at
this conference.

On the nonperturbative side, the success of the power correction analysis of event shapes
in electron-positron  annihilation and  DIS surprised just about
everyone.  The reasons for these successes, and their real extent, are not
yet fully understood.  In this regard we should be encouraged, and at
the same time more critical. 

Much more work needs to be done on the relation of power corrections to
perturbation theory in hadron-hadron scattering.  In this connection,
it will be important to sit back and think about
what new measurements and theoretical developments
will allow us to learn about quantum chromodynamics.

\subsection*{Acknowledgements}

I would like to thank the director of the Rencontres de Moriond, J.\ Tr\^an Thanh
V\^an and the program committee, particularly Dominique Schiff,
for their invitation and support.  I would also like to thank 
Gregory Korchemsky for many useful conversations, and for his encouragement and advice.
This work was supported in part by the National Science Foundation,
under grant PHY9722101

\end{document}